\documentclass[prb,
twocolumn,
superscriptaddress,showpacs,amsmath,amssymb]{revtex4}
\usepackage{amsfonts}
\usepackage{bm}
\usepackage{verbatim}

\usepackage{graphicx}

\begin{document}
\title{Orbital momentum of chiral superfluids and  spectral asymmetry of edge states}

\author{G.E.~Volovik}
\affiliation{O.V. Lounasmaa Laboratory, School of Science and Technology, Aalto University, Finland}
\affiliation{Landau Institute for Theoretical Physics, acad. Semyonov av., 1a, 142432,
Chernogolovka, Russia}

\date{\today}

\begin{abstract}
{This is comment to preprint arXiv:1409.7459
by
Y. Tada, Wenxing Nie and M. Oshikawa 
"Orbital angular momentum and spectral flow in two dimensional chiral superfluids",
\cite{Tada2014} where the effect of spectral flow along the edge states on the magnitude of the orbital angular momentum is discussed. The general conclusion of the preprint on the essential reduction of the angular momentum for the higher values of chirality, $|\nu|>1$, is confirmed. However, we show that if parity is violated, the reduction of the angular momentum takes place also for the $p$-wave superfluids with $|\nu|=1$.
}
\end{abstract}

\maketitle

\section{Introduction}

The problem of the orbital angular momentum of chiral fermionic superfluids
attracted new attention due to topological properties of these liquids and the related quantum anomalies. For the pair correlated chiral superfluids with angular momentum $\hbar \nu$ of the Cooper pair, the naturally expected value of the orbital angular momentum of the axially symmetric ground state is $L_z=\hbar \nu N/2$, see e.g. 
Ref. \cite{ShitadeKimura2014}. This expectation follows from the symmetry of the system:  the ground state many-body wave function $\Psi$ and the order parameter, which is proportional to $(p_x+ip_y)^\nu$, are invariant under the combined transformations:
\begin{equation}
\hat Q\Psi =0~~,~~\hat Q(p_x+ip_y)^\nu=0 ~~,~~ \hat Q =\hat L_z - \frac{\nu}{2}\hat N\,.
 \label{Qsymmetry}
\end{equation}
Here $\hat L_z$ is the operator of orbital $SO(2)$ rotations, and $\hat N$ is the operator of the global $U(1)$ symmetry transformations, which reflect the conservation law for particle number $N$.
The application of this $Q$-symmetry to the ground state gives
\begin{equation}
 L_z=\left<\hat L_z\right>=   \frac{\nu}{2}\hbar\left<\hat N\right>=\hbar \nu N/2 \,.
 \label{naive}
\end{equation} 

Another argument in favor of Eq.(\ref{naive}) is that it certainly works in the BEC limit, when the system represents the Bose-Einstein condensate (BEC) of $N/2$ molecules, each with angular momentum $\nu \hbar$. 
The adiabatic transformation from the BEC state to the BCS weak coupling regime does not change the value of the angular momentum, if the transformation preserves the axial symmetry. Thus the Eq.(\ref{naive}) should be valid in both regimes. The Eq.(\ref{naive}) has been supported by different calculations, see e.g.
Refs. \cite{McClureTakagi1979,StoneRoy2004,Sauls2011} and references therein.

However, this consideration does not take into account that the  system is finite: in the cylindrical vessel
the order parameter is inhomogeneous due to boundary conditions (texture), and the edge states on the boundary also intervene. During the transformation of the system from the BEC to the BCS regime, the phenomenon of spectral flow takes place. The accumulation of the angular momentum in this process may lead to the spectral asymmetry, which
essentially modifies $L_z$. It was demonstrated in Ref. \cite{Volovik1995} that in the three-dimensional $^3$He-A, where $\nu =1$, the spectral flow occurs through the Weyl points in the bulk fermionic spectrum. If parity is violated, the accumulation of angular momentum in the twisted texture takes place, and Eq.(\ref{naive}) fails. 

It was suggested  in Ref. \cite{Volovik1995}, that Eq.(\ref{naive}) may fail also due to the spectral flow through the gapless fermionic states, which  live either within the vortex cores or on the boundaries. In the recent paper Ref. \cite{Tada2014}  the effect of the edge states in the two-dimensional chiral superfluid has been discussed for general $\nu$. It was demonstrated that for $|\nu|=1$, the Eq.(\ref{naive}) is valid  in the considered geometry even in the BCS limit. However, for $|\nu|>1$ the orbital angular momentum is suppressed due to spectral asymmetry of the fermionic edge states. For the 2D case  in the BCS limit it was obtained
\begin{equation}
L_z=\frac{\hbar}{2} \nu \left(N - \frac{k_F^2}{\pi}V\right)~~,~~|\nu| >1\,.
 \label{general}
\end{equation}
Since in the BCS limit the 2D particle density $n=N/V$ practically coincides with $k_F^2/\pi$,
the suppression is crucial. It is interesting that Eq.(\ref{general}) corresponds to the so-called intrinsic angular momentum, which appears in the equations of the orbital dynamics.\cite{Volovik1975}  

The related issue of the reduction of the edge current in chiral superconductors has been discussed in Refs. \cite{HuangTaylorKallin2014,ScadiSimon2014}.
 
Here we apply the general approach of the spectral flow through the
gapless edge states to the situation, when the parity is broken or explicitly violated. Within our model we obtain  that equation (\ref{general}) for $|\nu| >1$ remains valid even if parity is violated.
However, for $|\nu|=1$,  the Eq.(\ref{naive}) is also modified, but the reduction of the angular momentum depends on the degree of the parity violation.  

\section{Accumulation of fermion charge due to spectral flow}

\begin{figure}
\includegraphics[width=1.0\linewidth]{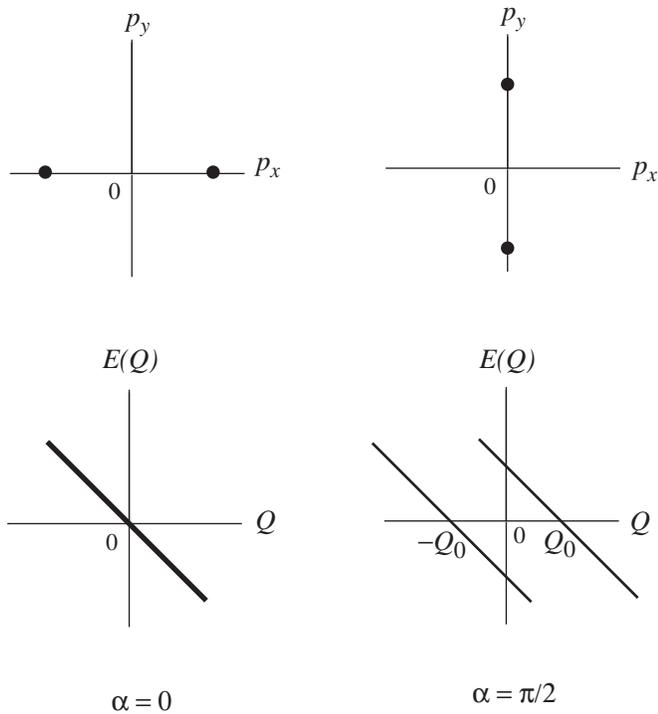}
\caption{
Fig.~\ref{N1}. Illustration of Weyl points in the phase space in the boundary layer and the corresponding anomalous branches of the spectrum of edge modes in the $p$-wave superfluid ($\nu = 1$).  ({\it top left}): two Weyl points for zero value of the twist parameter $\alpha=0$. The normal to the wall is along the axis $x$. Since the $p_y$ components of the Weyl points are zero, the anomalous branch $E(Q)$ in  ({\it bottom left}) crosses zero at $Q=0$. There is no accumulation of the angular momentum during the process of transformation from BEC to BCS, and thus the traditional equation for the total orbital angular momentum in Eq.(\ref{naive}) is not disturbed. ({\it top right}): two Weyl points in case of the twist parameter $\alpha=\pi/2$.   
Since the $p_y$ components of the Weyl points are now nonzero, $p_{y,a}=\pm p_F$, the anomalous branch $E(Q)$ splits into two branches, which cross zero at $Q=\pm p_F R$ ({\it bottom right}). This gives rise to the modification of Eq.(\ref{naive}). For the general twist parameter $\alpha$ the orbital momentum is given by Eq.(\ref{nu1}).
 }
 \label{N1}
\end{figure}

The edge states are characterized by the quantum number $Q$, which is the eigenvalue 
of the symmetry operator of the chiral superfluids in Eq.(\ref{Qsymmetry}).
The same symmetry characterizes the bound states inside the vortex with $\nu$ quanta of circulation in the non-chiral superfluids.\cite{SalomaaVolovik1987}
In both cases there are anomalous branches of spectrum of fermions, $E_a(Q)$, which cross zero energy, see Fig. \ref{N1} ({\it bottom}).
The number of the anomalous branches is proportional to $\nu$.\cite{Volovik1993} 
These branches connect the negative and positive energy levels and may realize the spectral flow of the angular momentum
under the deformation of the system from BEC to BCS. The latter occurs if the anomalous branch crosses zero at nonzero value of $Q$, i.e. if $E(Q_0)=0$ at $Q_0\neq 0$, see e.g. Fig. \ref{N1}({\it bottom right}). Then for each event of the level crossing the momentum $Q_0$ is added to or removed from the vacuum state. 
In the 2D case the total charge $\left<\hat Q\right>$ accumulated by the texture, by vortex or by the edge modes is given by  \cite{Volovik1995}
\begin{equation}
\left<\hat Q\right>=-\frac{1}{2}\sum_aQ_a^2 \,.
 \label{accumulation}
\end{equation}
Here $Q_a$ is the value of $Q$ at which the branch $E_a(Q)$ crosses zero. It is assumed here that all the anomalous branches have a negative slope, otherwise the algebraic sum must be used.

Let us find the charge $\left<\hat Q\right>$ accumulated by the edge states using a simple model of the boundary. We assume here that the characteristic length scale of variation of the order parameter 
near the wall is large compared to the coherence length. This allows us to use the semiclassical approximation. 

\subsection{Parity violated boundary conditions}

The effect of the boundary conditions on the angular momentum in superfluids with $\nu=1$ has been considered by Sauls 
\cite{Sauls2011}. It was shown that the specular boundary conditions do not violate  Eq.(\ref{naive}), while the diffusive boundary conditions essentially modify $L_z$. The possible reason for the latter is that the diffusive wall actually violates the axial symmetry of the system. That is why here we use the modification of the boundary conditions, which certainly obeys the axial symmetry, but violates parity.

 For that we choose the Hamiltonian near the wall of the container, which in  the semiclassical approximation has the form
\begin{equation}
{\cal H}= \tau_3\frac{k^2-k_F^2}{2m} +
 \Delta_\parallel(x)\tau_1     \cos(\nu\theta +\alpha)  + \Delta_\perp(x)\tau_2
\sin(\nu\theta  + \alpha).
 \label{QuasiclassicalHamiltonian}
\end{equation}
Here we consider the order parameter near the cylindrical wall close to the point  
$x=R$, $y=0$, where $R$ is the radius of the vessel; $k_x=k_F\cos\theta$; $k_y=k_F\sin\theta$. 
Far from the wall the components of the order parameter are equal,  $\Delta_\parallel=  \Delta_\perp\equiv \Delta$, 
which corresponds to the state $(p_x+ip_y)^\nu e^{i\alpha}$ for positive $\nu$ and  $(p_x-ip_y)^{|\nu|} e^{i\alpha}$ for the negative $\nu$. Near the wall the two components of the order parameter split. The phase $\alpha$ regulates the parity violation on the boundary. Such configuration of the order parameter near the wall is equivalent to the twisted texture considered in Ref. \cite{Volovik1995}, which led to reduction of the angular momentum of the texture.

\subsection{Weyl points in phase space and anomalous branches}

To study the edge states we extend the order parameter across the wall using  the  constraints  
$\Delta_\parallel(R-\tilde x)=-\Delta_\parallel(R+\tilde x)$ and 
$\Delta_\perp(R-\tilde x)=\Delta_\perp(R+\tilde x)$, and find the bound states in this extended potential. 
In this approach we double the number of the edge states, since only half of the solutions will satisfy the boundary conditions for the edge states. This, however, does not influence the result: we must keep in mind that the final result must be divided by factor 2. With $\alpha\neq 0$ this procedure generalizes the specular reflection boundary conditions to the situation, when parity is broken without violation of axial symmetry.

The semiclassical approximation allows us to use the topology in the phase space ($k_x,k_y,x)$. The Hamiltonian (\ref{QuasiclassicalHamiltonian}) as function of these three phase space variables contains $2\nu$ Weyl points, where all three terms in Eq. (\ref{QuasiclassicalHamiltonian}) are nullified. These points determined by equations $x=R$, $k=k_F$, $\sin(\nu\theta_a  + \alpha)=0$
are topologically protected. As a result, in the quantum-mechanical problem these points give rise to $2\nu$  anomalous branches 
$E_a(Q)$, which cross zero energy (clarify in Sec. 23 in Ref.  \cite{Volovik2003}, where it is shown how the anomalous branches $E_a(Q)$ emerge for fermions localized on vortices). The values   $Q_a$, where $E_a(Q_a)=0$, are $Q_a=k_{ya}R= k_F R\sin \theta_a$.

\subsection{$p$-wave superfluid}

Let us first consider the $p$-wave superfluid ($\nu=1$). There are two Weyl points in the phase space, with $\theta_1=-\alpha$ and $\theta_2=\pi-\alpha$.  Fig. \ref{N1} ({\it top left}) and Fig. \ref{N1} ({\it top right}) demonstrate the Weyl points for $\alpha=0$ and for $\alpha=\pi/2$ respectively. The Weyl points give two values of the charge $Q$, at which the energy of the edge modes $E_a(Q)$ is zero: $Q_{1,2}=  \pm k_F R\sin \alpha$. Using 
Eq.(\ref{accumulation}) one obtains for $\nu=1$ the following value of the total angular momentum:
\begin{equation}
L_z(\nu=1)=\frac{\hbar}{2}  N+\left<\hat Q\right>= \frac{\hbar}{2}  \left(N - 2\sin^2\alpha\frac{k_F^2}{\pi}V\right) \,.
 \label{nu1}
\end{equation}

The case $\alpha=0$ corresponds to the traditional specular boundary conditions without twist. In this case the anomalous branch crosses zero at $Q=0$ (see Fig. \ref{N1} {\it bottom left}). During the transformation from the BEC regime, where $k_F=0$, to the BCS regime, no accumulation of the angular momentum occurs. Thus the traditional equation (\ref{naive}) remains valid even in the BEC regime. 

For $\alpha\neq 0$, the twist splits the anomalous branch into two anomalous branches in Fig. \ref{N1} ({\it bottom right}). The spectral flow of the angular momentum along these branches during the transformation modifies the total orbital angular momentum of the system according to Eq.(\ref{nu1}).

\subsection{Superfluid with higher chirality}

\begin{figure}
\includegraphics[width=1.0\linewidth]{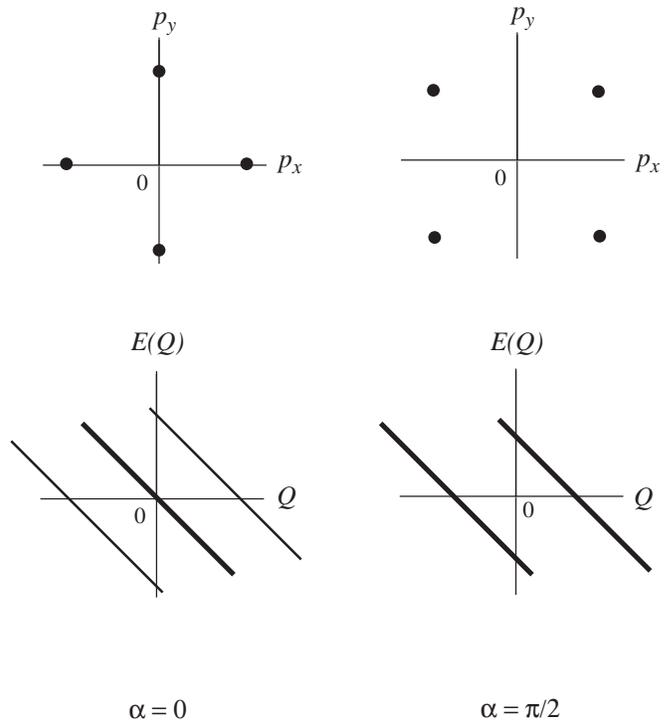}
\caption{
Fig.~\ref{N2}. The same as in Fig.~\ref{N1} but for  the $d$-wave superfluid with $\nu=2$.
Now there are four Weyl points in the boundary layer.  The $p_y$ components of the Weyl points are nonzero at least for two of them. That is why for all values of the parameter $\alpha$ there exist the edge modes whose spectrum crosses zero at $Q\neq 0$. This leads to the reduction of the orbital angular momentum according to Eq.(\ref{nu2}), which in our model does not depend on the twist parameter 
$\alpha$.
 }
 \label{N2}
\end{figure}

Now let us move to the liquids with higher chirality, $|\nu|>1$. For $\nu=2$ one has four Weyl  points in the phase space, which are obtained from each other by $\pi/2$ rotations, see Fig. \ref{N2} ({\it top left})
for $\alpha=0$ 
and Fig. \ref{N2} ({\it top right}) for $\alpha=\pi/2$. That is why there always exist the anomalous branches
which cross zero energy at $Q\neq 0$. Fig. \ref{N2} ({\it bottom left}) illustrates the branches for $\alpha=0$ 
and Fig. \ref{N2} ({\it bottom right}) --  the branches for $\alpha=\pi/2$. The latter corresponds to 
Fig. 2 in Ref. \cite{Tada2014}.

For general twist angle $\alpha$, the Weyl points are  at $2\theta_1=-\alpha$, $2\theta_{2,3}=\pm \pi-\alpha$, $2\theta_4=2\pi-\alpha$. Inserting the corresponding $Q_a$ to Eq.(\ref{accumulation}) one obtains that the total angular momentum does not depend on the twist $\alpha$:
\begin{equation}
L_z(\nu=2)= \hbar  N+\left<\hat Q\right>=  \hbar   \left(N -\frac{k_F^2}{\pi}V\right) \,.
 \label{nu2}
\end{equation}
Eq.(\ref{nu2}) agrees with Eq.(\ref{general}) obtained in Ref. \cite{Tada2014}. Consideration of the states with higher $|\nu|$ also supports the general Eq.(\ref{general}) with no dependence on twist.

\section{Conclusion}

Accumulation of the fermionic charge due to spectral flow gives us a simple tool for investigation of the problem of the orbital angular momentum in the axisymmetric chiral superfluids. It demonstrates that even without violation of the axial symmetry the angular momentum can be essentially reduced compared with its natural value $\nu \hbar N/2$. The spectral flow occurs through the gap nodes in the spectrum, which are either the Weyl points in the bulk chiral liquid or the Weyl points in the mixed phase space 
describing fermions living in the vortex cores or on the boundary of the system.
This is one of the realizations of the chiral anomaly in condensed matter.

Due to the spectral flow effect the magnitude of the total orbital angular  momentum looses its universality.
In particular, in the $p$-wave superfluid it is determined by the degree of the parity violation, as demonstrated in 
Eq.(\ref{nu1}), which  depends on the twist parameter $\alpha$. 
For the higher $|\nu|>1$, the reduction of the angular momentum in Eq.(\ref{general}) does not depend on 
the parity violation parameter. However, this can be the artifact of the models used.

\section*{\small Acknowledgements}
I acknowledge the financial support by the Academy of
Finland through its LTQ CoE grant (project $\#$250280).

\end{document}